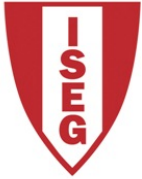

# MASTER
## ECONOMICS

# FINAL MASTERS WORK
## DISSERTATION

MODELING STOCK MARKETS THROUGH THE RECONSTRUCTION OF MARKET PROCESSES

JOÃO PEDRO RODRIGUES DO CARMO

OCTOBER — 2017

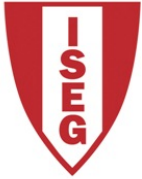

# Master in
## Economics

# Final Masters Work
## Dissertation

Modeling Stock Markets Through The Reconstruction Of Market Processes

João Pedro Rodrigues do Carmo

**Supervisor:**
Tanya Vianna de Araújo

October — 2017

**Abstract**

There are two possible ways of interpreting the seemingly stochastic nature of financial markets: the Efficient Market Hypothesis (EMH) and a set of stylized facts that drive the behavior of the markets. We show evidence for some of the stylized facts such as memory-like phenomena in price volatility in the short term, a power-law behavior and non-linear dependencies on the returns.

Given this, we construct a model of the market using Markov chains. Then, we develop an algorithm that can be generalized for any N-symbol alphabet and K-length Markov chain. Using this tool, we are able to show that it's, at least, always better than a completely random model such as a Random Walk. The code is written in MATLAB and maintained in GitHub.

**Table of Contents**

**Acknowledgements**

First and foremost, I'd like to dedicate this work to my parents who have always shown me immense support and whom without this wouldn't even be possible.

I'd like to give a sincere acknowledgement to my supervisor, for her incredible support throughout the entire time this work has taken. I'd like to thank her for pushing me in the right direction, for providing an exciting subject to work on and for all the patience.

# 1. Introduction

The financial markets are seemingly stochastic, but there are two major and different ways of interpreting such property. On one hand, we can believe in a set of stylized facts — that is, an empirical finding that is true more often than not — about the markets and their behavior. On the other hand, we can believe in the Efficient Market Hypothesis (EMH) and that it's impossible to beat financial markets due to their absolute knowledge of available information.

The literature is vast and support is widely distributed among each perspective. We can point some popular stylized facts identified (Cont, 2001) about financial assets such as the existence of memory-like phenomena in price volatility, power-law behavior in returns, correlations between returns of distinct companies and non-linear dependencies on the returns.

The EMH can assume the form of one of its three variants (Fama, Efficient Capital Markets: A Review of Theory and Empirical Work, 1970): weak, semi-strong and strong, but all share the claim that market efficiency causes the prices on traded assets to incorporate all publicly available past information. As a result, the assets are traded at their fair value and it's impossible to take advantage of market flaws either through undervalued assets, inflated prices or timing mechanisms.

There has been work made that contributes to both ends of the question, supporting the apparent efficiency of markets under linear statistical tests and failure to outperform the markets by practitioners while providing evidence for non-linear forecasting methods to achieve above average returns (Sewell, 2012).

This and the existence of stylized facts seem to contradict the lack of structure in financial markets. In fact, we can think of them as complex systems altogether in that they meet most, if not all, of the required criteria: memory and feedback, non-stationarity, a multitude of interacting agents with adaptation and evolution exhibiting extreme behavior — remaining far from equilibrium, being a single realization and an open system with the environment (Johnson, Jefferies, & Hui, 2003).

The financial markets can be seen as an example of a complex adaptive system and are certainly one of the most complex structures known with a very unique and distinct property: their building blocks — the agents — are intelligent beings. Investors are quick to react and are always searching for the best possible outcome.

In order to achieve better than average risk-adjusted returns, the agents try a variety of tools that they have at their disposal. Borrowing from many different disciplines, we have available a multitude of tools that have been designed to study the structure of complex systems whose emergent behavior cannot be reduced to the study of its parts separately. Traditionally, this has been applied in the natural physical sciences, but has since been adopted by many areas such as biology, sociology and, of course, economics.

One of the available tools for the study of stochastic processes is modelling the underlying dynamics as a Markov chain. This can be applied to Markov processes which are stochastic processes satisfying the Markovian — or *memoryless* — property. That is, the future states depend on the history (of states) only through the current state (Serfozo, 2009). A Markov process becomes a Markov chain if it has a discrete state space. Continuous processes can also be reduced to Markov chains if we can describe the time as a countable state space.

The purpose of modelling a process is not only to describe its past and understand its current behavior, but also to try and get some insight into its possible future path. One can argue that for as long as markets have existed, there have been those who tried to beat them. Though arbitrage, insider information and other methods may be valid for some markets, they aren't for the financial markets as such advantages are so readily resolved that they can be considered practically inexistent. One must then try to get advantage in a different way, for example, by trying to predict the future price of a given asset and thus contrive a strategy to achieve a profit.

Admittedly, the stock market does not meet the criteria of complete independence of present price movements from past ones, but these influences are arguably so small that they fail to be useful to an investor (Malkiel, 1973). This fact renders buy-and-hold strategies useless. Thus, one can argue in favor of the Random Walk Hypothesis (RWH) and that price evolution is due to an unpredictable random walk consistent with the EMH.

The prediction methods can be characterized as one of three larger categories that may overlap: fundamental analysis, technical analysis and data mining technologies. Fundamental analysis is concerned with the intrinsic value of a stock and studies that intrinsic value, including the performance of the company behind the stock and the overall economy. Technical — or chart — analysis, on the other hand, evaluates market statistics and tries to identify patterns in the data. Finally, data mining technologies borrow the power of computers and techniques from other fields such as Artificial Neural Networks (ANN) and Genetic Algorithms (GA). We can include the tools from complex theory in this set, such as the Markov Chain model.

In a 1993 letter to the Shareholders of Berkshire Hathaway Inc., Warren Buffet quoted the American economist Ben Graham as "In the short-run, the market is a voting machine — reflecting a voter-registration test that requires only money, not intelligence or emotional stability — but in the long-run, the market is a weighing machine" (Buffett, 1993). What he means is that emotions control the short-run while a company's assets and profits control the long-run. We can see one example of this when Twitter mood can be used to predict up and down movements in the closing values of the Dow Jones Industrial Average (DJIA) with an accuracy of 86.7% (Bollen, Mao, & Zeng, 2011). Even though the short-run is very volatile, the long-run follows a more stable path and, at least in comparison, more predictable.

The main source of both information and motivation for this work is the paper presented in reference (Vilela Mendes, Lima, & Araújo, 2002) in which the authors develop a market reconstruction procedure based on the market fluctuations. They conclude that it's a short-memory process with a small long-memory component which suggests that *chains with complete connections and summable decays* are an appropriate model for it. I intend to further validate some of their results using more data points and an improved algorithm that should not only be general, but also readily available and easy to adapt to different study cases.

More specifically, my main contribution relies in the following:

1. Confirmation of previous results
2. Development of a general algorithm for N-symbol and K-length Markov chains
3. Forecasting real prices from the forecasted returns

Markov chains have been shown to reproduce most of the stylized facts that we know about daily series of returns (Bulla & Bulla, 2006). They have been used as well in an attempt to capture more accurately the evolution of a risky asset (Xi & Mamon, 2011), to study the high frequency price dynamics of traded stocks (D'Amico & Petroni, 2012) and to predict loan defaults on credit (Vojtekova, 2013). In (Stadnik, 2014) they attempt to find an appropriate mathematical description of the financial market distributions based on Markov chains and in (Xi, Peng, & Qin, 2016) they make use of the model under Monte Carlo to estimate the leverage effect in financial time series.

In this work, I will present a general algorithm to calculate the Markov (or transition) matrices not only in two dimensions ($i_1 \times i_2$) as is the case when only the previous day is considered in the calculations, but also in $n$ dimensions which will produce Markov tensors instead with size ($i_1 \times i_2 \times ... \times i_n$) as in the case when lengths of $K$ previous days are considered. The algorithm was written in MATLAB, stored and maintained in an online public repository that also provides version control (GitHub). This is also fully documented for an easier understanding of the written code. The permanent URL is:

https://github.com/joaocarmo/market-reconstruction/

The data source is the historical prices obtained from the Yahoo! Finance website for different companies and indexes. This choice allows for easy replications and adaptations of the current work by anyone willing as both the data source and the code are provided without barriers and on demand to the public. I intend to create and provide a framework that is available for easy replication and adaptable to future works while providing a replication study of previous research.

The results will be presented, whenever possible, compared with the ones obtained in reference (Vilela Mendes, Lima, & Araújo, 2002). Not all the results of that paper are replicated in this work and some of the results here have not been tried previously.

In the following chapters, I will present a brief look into the stochastic nature of the financial markets and the basics of Markov chains, a detailed description of the algorithm implemented and the results obtained.

## 2. The stochastic nature of the financial markets

Developing and testing models for the behavior of financial markets has been the interest of many economists, mathematicians and even physicists for years. In spite of their advances, actual market practitioners employ mostly one of two approaches available to predict stock prices: chartist theories and the intrinsic value analysis.

The chartist theories are based on the assumption that there are price patterns in the history of a stock price that can be analyzed. Identifying such patterns in the past allows the recognition of situations in the future that are likely to occur. Dow theory is an example of this method (Malkiel, 1973).

The intrinsic value analysis, on the other hand, assumes that a stock always has an intrinsic value or equilibrium price which depends on the company itself and the economy as a whole. By studying the micro and macroeconomic variables surrounding the stock, an analyst can infer how much the price of the stocks differs from its intrinsic value.

Both approaches are contested by (Fama, Random Walks in Stock Market Prices, 1965) and do not have academic relevance. Other theories such as the Theory of Random Walks have stronger support from the empirical evidence. This is a theory in which prices evolve according to a random walk which renders its prediction attempts useless. A random walk is a stochastic process and it describes a path constructed from a succession of random steps through a mathematical relationship in which the next value depends on the previous one. Many random walks have a corresponding representation as Markov chains.

## 2.1. The market as a random walk

The RWH is a popular theory stating that stock market prices are unable to be predicted and thus evolve according to a random walk. This is a consequence of the EMH in which future prices cannot be forecasted based on past performance. In order words, it has the Markov property or *memorylessness* and any time series which satisfies the Markov property is also a Markov process.

Thus, we can say that random walks are one example of a Markov process. This idea was originally proposed in (Bachelier, 1900).

We can construct a simple one-dimensional random walk to model the stock market by following these rules:

1. Every day, the price is decided by a coin toss
2. The coin is unbiased, so the chances of both heads and tails are equal
3. If it's heads, the price goes up by one unit
4. If it's tails, the price goes down by one unit

Let $P(u)$ be the probability of flipping *heads* and $P(d)$ the probability of flipping *tails*, we have that $P(u) = P(d) = 0.5$, $P(u) + P(d) = 1$ and the price evolution is given by eq. (1).

$$p_t = \begin{cases} p_{t-1} + 1, & heads \\ p_{t-1} - 1, & tails \end{cases} \quad (1)$$

We can see that the expected value is constant $E[p_t] = P(u)(p_{t-1} + 1) + P(d)(p_{t-1} - 1) = p_{t-1}$ and the real value just circles back and forth around the starting value $p_0$. The probability distribution for $p_t$ follows a normal distribution with $\mu = p_0$ and $\sigma = 0.5\sqrt{t}$, where $t$ is the number of days.

In figure 1, we can clearly see that as we increase the number of steps, the graphical representation of $p(t)$ grows in similarity with the real stock market.

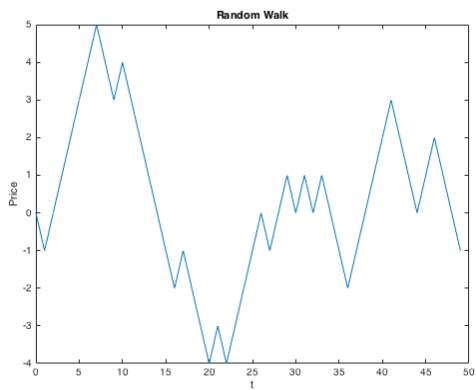
(a)

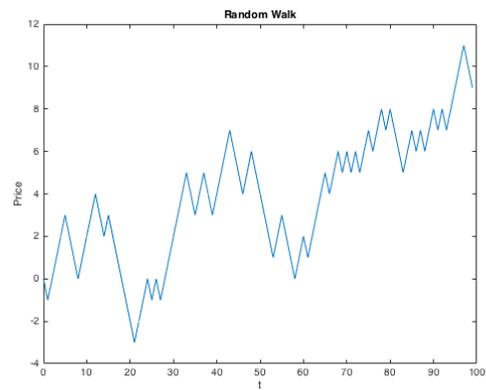
(b)

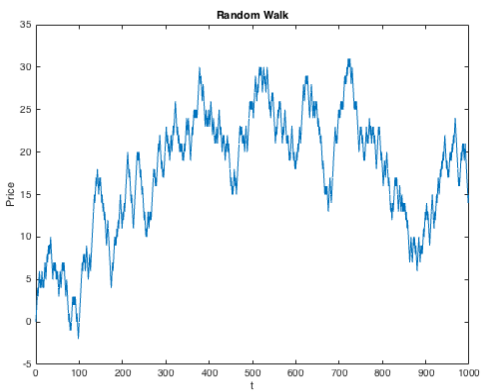
(c)

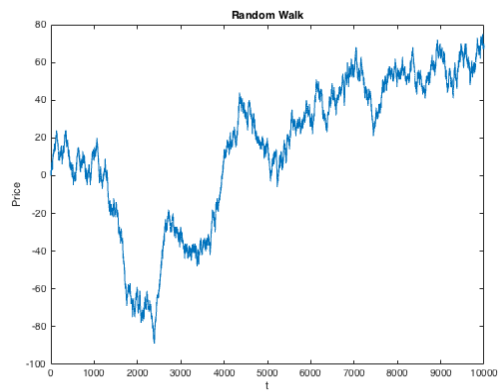
(d)

Figure 1: A graphical representation of a random walk process with (a) 50, (b) 100, (c) 1 000 and (d) 10 000 steps

## 2.2. The random walk as a Markov chain

We can model the previous example as a Markov chain using only the following simple rules:

1. Each day the price must change
2. The price has equal probability of going up or down
3. The change in price is one unit

The price at any given time depends only on the price of the previous period and a certain probability. Given the current state of the system, it must move to a new state and it has equal probability of going to one of two possible new states. In one, the price goes up and in the other the price goes down (figure 2).

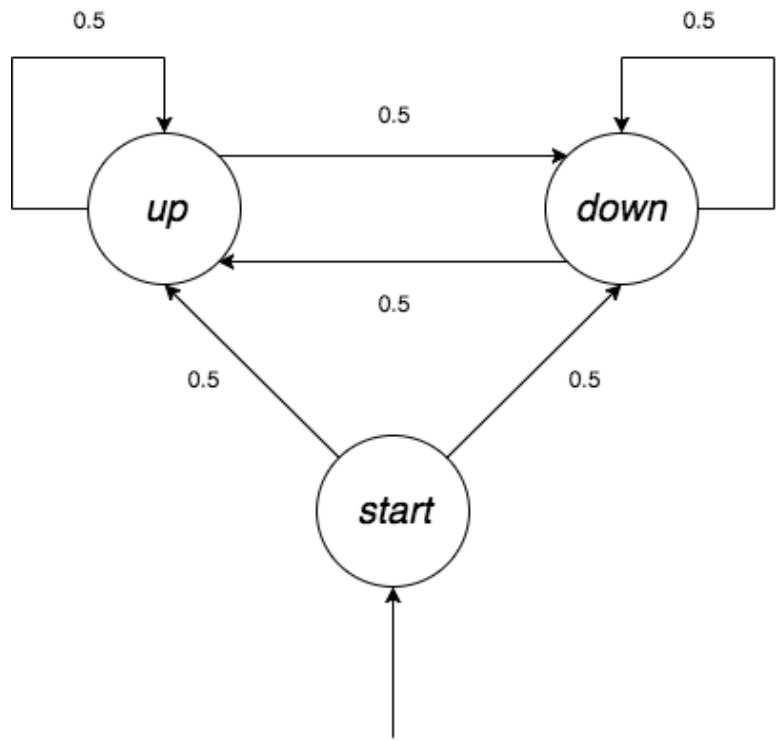

Figure 2: Transition diagram for a coin toss as a Markov chain

Let $P(i|j) = p_{i,j}$ be the probability of going from state $j$ to the state $i$, where $i, j \in S = \{start, up, down\} = \{0, u, d\}$. We also know that $p_{i,j} = 0.5$ when $i \neq 0$ because there is an equal probability of going in either direction regardless of what came before and you can't go back to the start. We can write the transition matrix as:

$$P = \begin{bmatrix} P(d|d) & P(d|0) & P(d|u) \\ P(0|d) & P(0|0) & P(0|u) \\ P(u|d) & P(u|0) & P(u|u) \end{bmatrix} = \begin{bmatrix} 0.5 & 0 & 0.5 \\ 0.5 & 0 & 0.5 \\ 0.5 & 0 & 0.5 \end{bmatrix}$$

And it follows that the transition probability from any state $j$ to any other state $i$ is 1 or $\sum_{i=0}^{S} p_{i,j} = 1$.

### 2.3. What are Markov chains?

Markov chains, named after the Russian mathematician Andrei Markov, are mathematical systems that loop between all possible states within a *state space* $S = \{s_1, s_2, \ldots, s_n\}$. From each given state $s_i$, there is a well-defined probability of jumping into a different state $s_j$. These probabilities can be arranged into a matrix called the *transition matrix P* (or stochastic matrix).

We can construct the transition matrix by adding each state within the state space as a row and as a column. This means that each element $(i, j)$ of the matrix describes the probability of transitioning from the column state $i$ to the row state $j$. In other words, we have the conditional probability $P(i|j) = p_{i,j}$. This will be a square matrix with dimensions $S \times S$.

More formally, a Markov chain is a set of random variables $\{X_t\}$ with $t \in \mathbb{N}_0$ in which the future is independent from the past, i.e. $P(X_t|X_0, X_1, \ldots, X_{t-1}) = P(X_t|X_{t-1})$. This is called the *memorylessness* property.

### 2.4. Modeling the market as a Markov chain

Considering the approach in (Vilela Mendes, Lima, & Araújo, 2002), we will model the market as a Markov chain using the simplest case which can later be easily expanded as needed. We start by considering a 3-symbol alphabet $\Sigma = \{d, 0, u\} = \{-1, 0, 1\}$ and consider the 1-day return rate $r(t, 1)$ given by eq. (2) for the prices $p(t)$ when we have $n = 1$.

$$r(t, n) = \log p(t + n) - \log p(t) \quad (2)$$

Let $\langle r(t, n) \rangle$ be the average value of $r(t, n)$ and $\sigma_n$ be the n-day standard deviation given by eq. (3) and eq. (4), respectively.

$$\langle r(t, n) \rangle = \frac{1}{N} \sum_{t=0}^{N} r(t, n) \quad (3)$$

$$\sigma_n = \sqrt{\langle (r(t, n))^2 - \langle r(t, n) \rangle^2 \rangle} \quad (4)$$

We can construct a chain $S$ of discrete events $s_i \in \Sigma$ based on the function (5), where $r = r(t, 1)$ is the 1-day return rate for a given moment $t$.

$$s_i(r) = \begin{cases} -1, & -\sigma > (r - \langle r \rangle) \\ 0, & \sigma \geq (r - \langle r \rangle) \geq -\sigma \\ 1, & (r - \langle r \rangle) > \sigma \end{cases} \quad (5)$$

Using this coding, we can translate the real valued function $r(t)$ into a discrete sequence of states $S = \{s_0, s_1, \ldots, s_N\}$ using the alphabet Σ: price goes down (-1), price stays the same (0) and price goes up (1). Notice the sequence below:

$$S_0 = \{0, 0, 1, 1, 1, -1, 0, 0, 0, 1, 1, 1, -1, 1, 0, -1, -1, \ldots\}$$

The sequence $S_0$ means that, for $r_0(t)$, the price stayed the same (within a $2\sigma$ margin) for the first two days ($s_0 = s_1 = 0$), then it went up during the next three days in a row ($s_2 = s_3 = s_4 = 1$) and then it came down ($s_5 = -1$) before stabilizing during the following three days ($s_6 = s_7 = s_8 = 0$), and so on.

The transition matrix for this chain will be constructed using the possible states within the state space (which is the alphabet $\Sigma$):

$$P = \begin{bmatrix} P(d|d) & P(d|0) & P(d|u) \\ P(0|d) & P(0|0) & P(0|u) \\ P(u|d) & P(u|0) & P(u|u) \end{bmatrix} = \begin{bmatrix} P(-1|-1) & P(-1|0) & P(-1|1) \\ P(0|-1) & P(0|0) & P(0|1) \\ P(1|-1) & P(1|0) & P(1|1) \end{bmatrix}$$

Where $P(i|j) = p_{i,j}$ is the conditional probability of going from the state $j$ to $i$ and is calculated based on the observed sequence $S$ according to eq. (6) with $n_{i,j}$ being the number of times the sequence $\{i, j\}$ has occurred in the $N - 1$ possible occurrences of a pair.

$$P(i|j) = \frac{n_{i,j}}{N - 1} \tag{6}$$

## 2.5. Forecasting the next value

We can start the reconstruction process after we've calculated successfully the transition matrix $P$ and we'll store the information in a new sequence $S^* = \{s_0^*, s_1^*, \ldots, s_N^*\}$. To determine each $s_{i+1}^*$ we need to follow the following steps:

1. Set $s_0^* = s_0$
2. Read the value $s_i$ from the original sequence $S$
3. Select its corresponding column $P(*|s_i)$
4. Throw a uniformly distributed random number $\varepsilon \in [0, 1]$

5. Set $s^*_{i+1} = f(\varepsilon, s_i)$

Where the function $f(x, s_1)$ is built using $P(*|s_i)$ according to eq. (7).

$$f(x, s_i) = \begin{cases} -1, & x < P(-1|s_i) \\ 0, & P(-1|s_i) \leq x < P(0|s_i) \\ 1, & x \geq P(1|s_i) \end{cases} \qquad (7)$$

## 2.6. Calculating the error

After obtaining the new sequence $S^*$, we can compare it with the original $S$ sequence and calculate the error according to eq. (8).

$$err = \sqrt{\sum_{t=0}^{N}(s_t - s^*_t)^2} \qquad (8)$$

## 2.7. Using an N-symbol alphabet

We can increase the alphabet $\Sigma$ in order to obtain a more fine-grained specification of the movements in $p(t)$. This can be seen as analogous to converting an analog signal to a digital one. We can imagine a sinusoidal wave, such as an electric signal coming from a microphone, which gets converted to a binary sequence. We could record only up and down movements or we could try and record every value in between. This increase in detail allows for a better reconstruction of the signal later on.

For a symmetric alphabet centered around 0, we need an odd total number of symbols available. We can then use the range:

$$\Sigma = \{-\beta, -\beta + 1, \ldots, -1, 0, 1, \ldots, \beta - 1, \beta\}$$

Where $\beta = \frac{N-1}{2}$ and the chain $S$ of discrete events $s_i \in \Sigma$ is constructed using eq. (9).

$$s_i(r) = \begin{cases} -\beta, & -\sigma > (r - \langle r \rangle) \\ \vdots & \\ 0, & \frac{\sigma}{\beta} \geq (r - \langle r \rangle) \geq -\frac{\sigma}{\beta} \\ \vdots & \\ \beta, & (r - \langle r \rangle) > \sigma \end{cases} \qquad (9)$$

The transition matrix $P$ will always be a square $N \times N$ matrix:

$$P = \begin{bmatrix} P(-\beta|-\beta) & P(-\beta|-\beta+1) & \cdots & P(-\beta|\beta-1) & P(-\beta|\beta) \\ P(-\beta+1|-\beta) & P(-\beta+1|-\beta+1) & \cdots & P(-\beta+1|\beta-1) & P(-\beta+1|\beta) \\ \vdots & \vdots & \ddots & \vdots & \vdots \\ P(\beta-1|-\beta) & P(\beta-1|-\beta+1) & \cdots & P(\beta-1|\beta-1) & P(\beta-1|\beta) \\ P(\beta|-\beta) & P(\beta|-\beta+1) & \cdots & P(\beta|\beta-1) & P(\beta|\beta) \end{bmatrix}$$

## 2.8. Using a K-length Markov chain

In the previous pages, we've been always assuming that the adjacent possible states to any given present state are always a function of the immediate past value. That is, given that the previous value is $u$, then all the possible states are given by the column $u$ in the transition matrix. We've kept $u$ as a single value up until now — a 1-length Markov chain — but that doesn't have to be so. In fact, we can construct any K-length Markov chain process as long as the data permits it — that is, the sequence exists.

In figure 3 it's represented a $K = 1$ length Markov chain with a 3-symbol alphabet $\Sigma = \{-1, 0, 1\}$ and its respective transition probabilities that make up the transition matrix $P$.

In a very general way, we could construct any K-length Markov chain with a N-symbol alphabet. But we'll construct a simpler 2-length Markov chain with the same 3-symbol alphabet we've used previously and then generalize it to any length $K$.

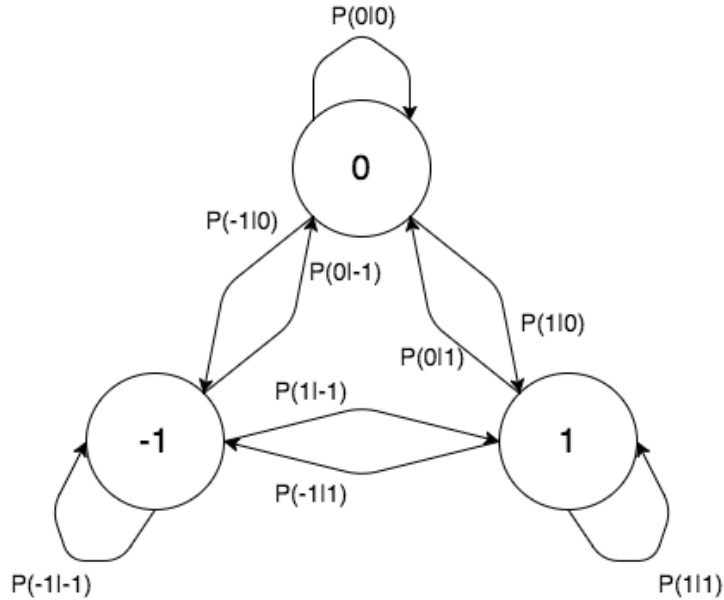

Figure 3: A three state Markov chain

Under these conditions, our state space is now:

$$S = \{-1-1, -10, 0-1, -11, 1-1, 00, 01, 10, 11\}$$

Where $\{-1-1\}$ means two consecutive days where the price went down and so on. This should no longer be described by a transition matrix, but by $3 \times 3 \times 3$ transition tensor $T$:

$$T_{i,j,-1} = \begin{bmatrix} P(-1|-1-1) & P(-1|0-1) & P(-1|1-1) \\ P(0|-1-1) & P(0|0-1) & P(0|1-1) \\ P(1|-1-1) & P(1|0-1) & P(1|1-1) \end{bmatrix}$$

$$T_{i,j,0} = \begin{bmatrix} P(-1|-10) & P(-1|00) & P(-1|10) \\ P(0|-10) & P(0|00) & P(0|10) \\ P(1|-10) & P(1|00) & P(1|10) \end{bmatrix}$$

$$T_{i,j,1} = \begin{bmatrix} P(-1|-11) & P(-1|01) & P(-1|11) \\ P(0|-11) & P(0|01) & P(0|11) \\ P(1|-11) & P(1|01) & P(1|11) \end{bmatrix}$$

Where $P(i|jk) = T_{i,j,k}$ is the conditional probability of getting $i$ given $\{jk\} \in S$ and $i,j,k \in \Sigma$. So, in general, we can build any $N^{K+1}$ transition tensor $T$ where each cell is given by eq. (10).

$$T_{i_1,i_2,\ldots,i_{K+1}} = P(i_1|i_2 \ldots i_{K+1}) \qquad (10)$$

And we know that the sequence $\{i_2 \ldots i_{K+1}\} \in S$ and $i_z \in \Sigma$, the N-symbol alphabet. We should notice that as K grows larger, the number of occurrences of the sequence $\{i_2 \ldots i_{K+1}\}$ in the data diminishes and it may happen that, for some K, the sequence does not occur at all in which case we have that $P(*|i_2 \ldots i_{K+1})$ is 0. If this happens, we shall consider the first sequence $\{i_2 \ldots i_z\}$ with $z < K + 1$ that satisfies $P(*|i_2 \ldots i_z) > 0$ starting with $z = K$ and reducing one at a time.

## 3. Empirical Data

The data source for this work is the historical data provided by Yahoo! Finance. We used the adjusted close price adjusted for both dividends and splits. For the main text, we chose the prices for *International Business Machines Corporation* (IBM) — NYSE. For the appendix, we chose *Facebook, Inc.* (FB) — NasdaqGS and *Alphabet Inc.* (GOOG) — NasdaqGS.

The data is accessible through https://finance.yahoo.com, searching for the company in question and then selecting "Historical Data." We can download the raw data in a comma-separated file (CSV) that contains 7 columns of information. We use the first one for the date and the sixth for the adjusted close price.

**4. Testing the stationarity of the process**

A stationary process is stochastic process which has the property that the mean, variance and autocorrelation structure do not change over time. White noise is a good example of a stationary process. Stationarity, for our purposes, means a seemingly flat series without trend and periodic fluctuations and with constant variance and autocorrelation structure over time.

We will use the most recent data from Yahoo! Finance and MATLAB to try and reproduce the stationarity tests in (Vilela Mendes, Lima, & Araújo, 2002) using the same 1-day market fluctuation data. In order to make a successful application of the statistical mechanics tools to these signals, two conditions should be fulfilled:

1. The process generating the data has some underlying stationarity as defined previously
2. The presented time sequence is a typical sample of the process

We shall process the data in a way that is consistent with the first condition while the analysis of different data sources (different stocks, currencies or indexes) should suffice for the second. This is easily done with the MATLAB code supplied in this work and because the results are everywhere similar, we will focus our analysis only in the International Business Machines (IBM) stock's daily closing price. Results for other data sources are available in the appendix.

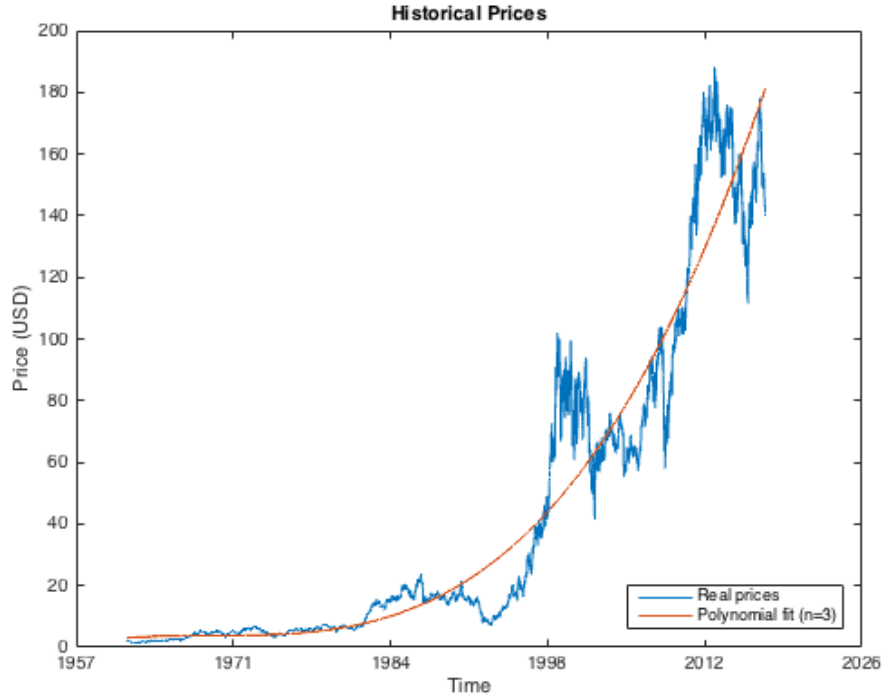

Figure 4: IBM's stock daily closing price over time and polynomial fit

In figure 4 we can see the historical price evolution $p(t)$ for IBM's stock and its corresponding 3rd degree polynomial fit $q(t)$. In figure 5a, we have removed the trend by applying eq. (11). We can see that even though the trend is gone, we still need to rescale the data. The rescaling is done by applying eq. (12) and we obtain the results in figure 5b. Now, the data looks much more like figure 1.

$$\bar{p}(t) = p(t) - q(t) \tag{11}$$

$$x(t) = \bar{p}(t)\frac{\langle p(t) \rangle}{q(t)} \tag{12}$$

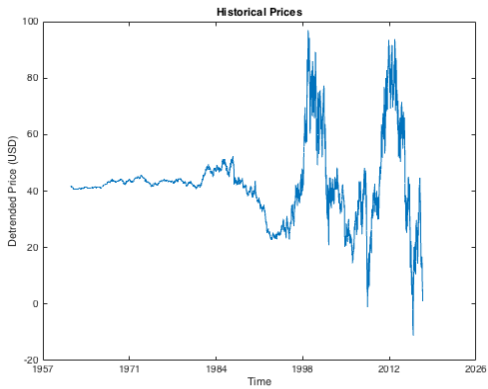 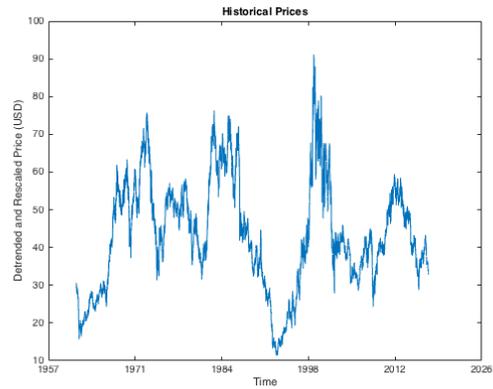

(a)                              (b)

Figure 5: The (a) detrended and (b) rescaled prices

In figure 6, we can see the 1-day returns over time according to eq. (2) and the dynamics of 1-day plotted as $r(t, 1)$ vs. $r(t + 1, 1)$ returns as a central core of small fluctuations with an outer aura of larger fluctuations.

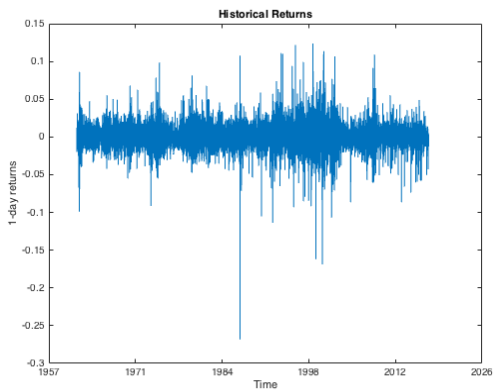 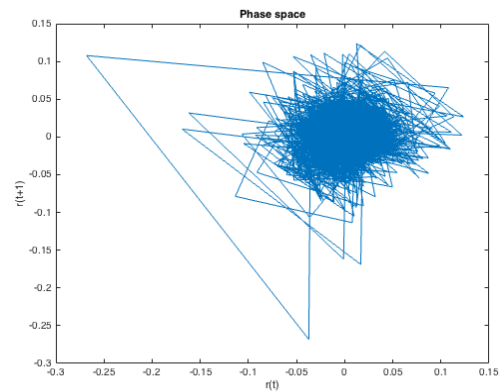

(a)                              (b)

Figure 6: The (a) 1-day returns and (b) the dynamics of 1-day returns

The figure 7 shows a strong variation in time of the 10-day window volatility as defined in eq. (4) which seems to indicate that the process is not locally stationary, but may be asymptotically stationary.

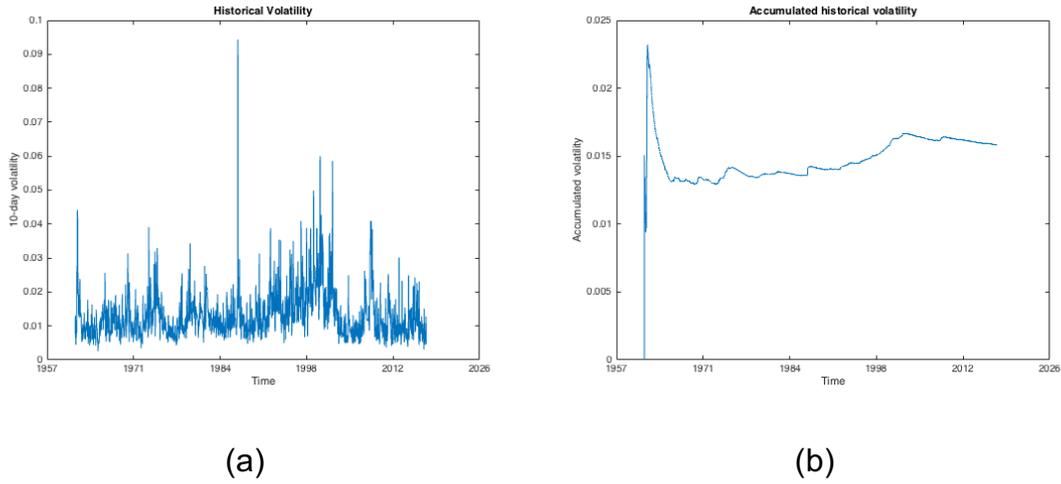

(a)          (b)

Figure 7: A (a) 10-day sliding window of the historical price volatility and (b) the accumulated volatility

Some other important statistical indicators can also be computed:

i. The maximum of $r(t,n)$ over $t$

$$\delta(n) = \max_t \{r(t,n)\} \qquad (13)$$

ii. The moments of the distribution of $|r(t,n)|$

$$S_q(n) = \langle |r(t,n)|^q \rangle \qquad (14)$$

iii. Within a certain range, satisfying

$$S_q(n) \sim n^{\chi(q)} \qquad (15)$$

In the figures 8 and 9 we have represented $\delta(n)$, $S_q(n)$ and $\chi(q)$ and the same conclusions as in (Vilela Mendes, Lima, & Araújo, 2002) can be observed. We can find an expression for $\chi(q)$ by rearranging eq. (15) and filling it with eq. (14) to obtain $\chi(q) = q \frac{\log|r(t,n)|}{\log n}$.

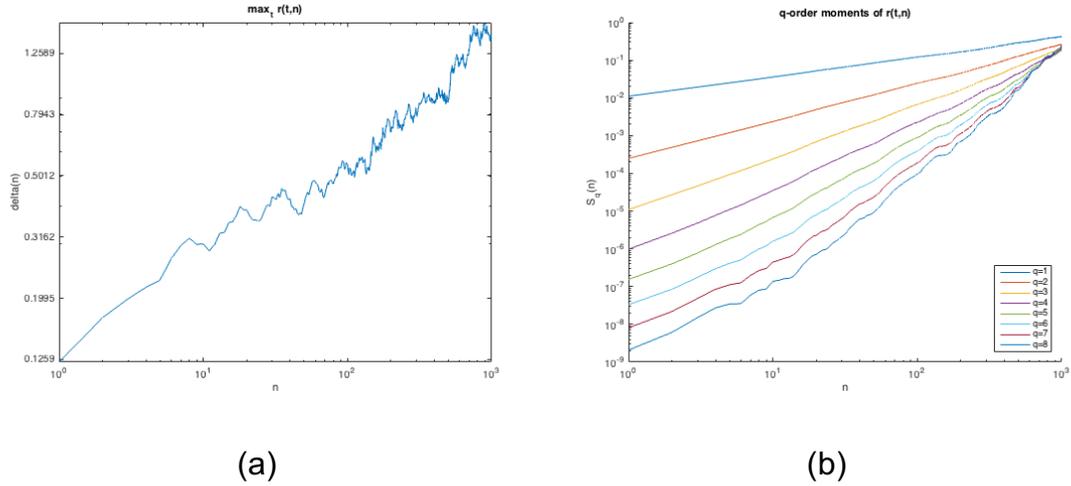

(a)           (b)

Figure 8: The plots of (a) the maximum $\delta(n)$ with $n$ from 1 to 1 000 and (b) the moments $S_q(n)$ with $q$ from 1 to 8 from top to bottom

In a nutshell, we can conclude from these statistical indicators that:

a. $\delta(n)$ is log-concave and probably asymptotically constant for large $r$

b. $S_q(n)$ is an increasing log-concave function allowing a power law approximation

c. $\chi(q)$ is an increasing concave function of $q$

As we can see from figure 8.a, by increasing the time range considered we also get an increase in the maximum value for the returns in that given time range and the tendency is log-concave. From figure 8.b, we see that the lines on the graph given by eq. (14) could be approximated through eq. (15).

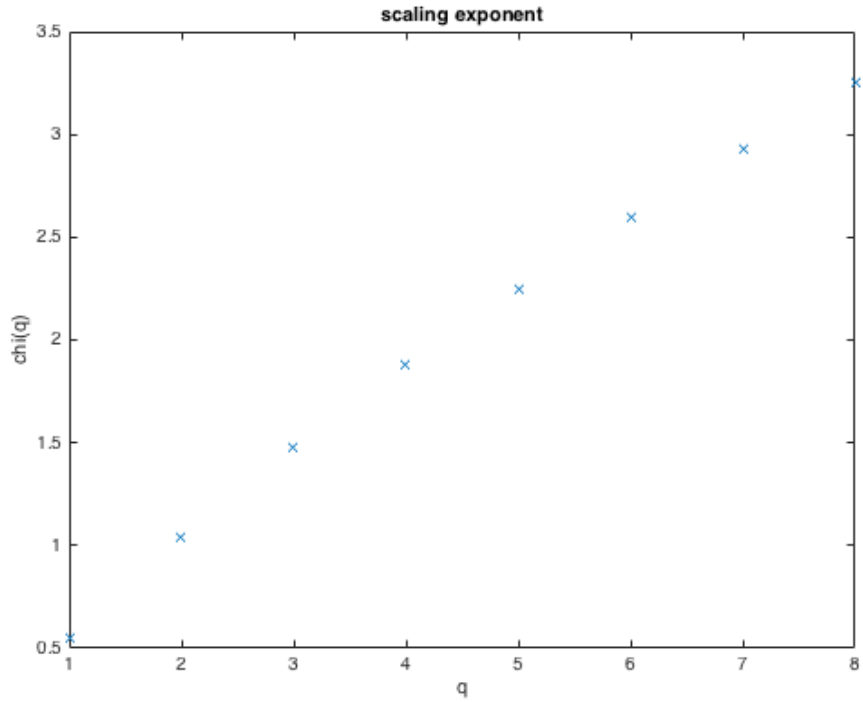

Figure 9: The scaling exponent $\chi(q)$

These conclusions are shared by turbulence data drawing a similarity between hydrodynamic turbulence and market fluctuations although with different numerical values. We should also note that $\chi(1) \approx 0.5$ which makes the signal uncorrelated for $n \geq 2$. We can further confirm this through the correlation between 1-day returns and its absolute value with eq. (16) and eq. (17).

$$C(r(t,1),T) = \langle r(t+T,1)r(t,1) \rangle \quad (16)$$

$$C(|r(t,1)|,T) = \langle |r(t+T,1)||r(t,1)| \rangle \quad (17)$$

As we can see in figure 10, the returns are uncorrelated for $T \geq 2$ with the correlation function remaining at the noise level.

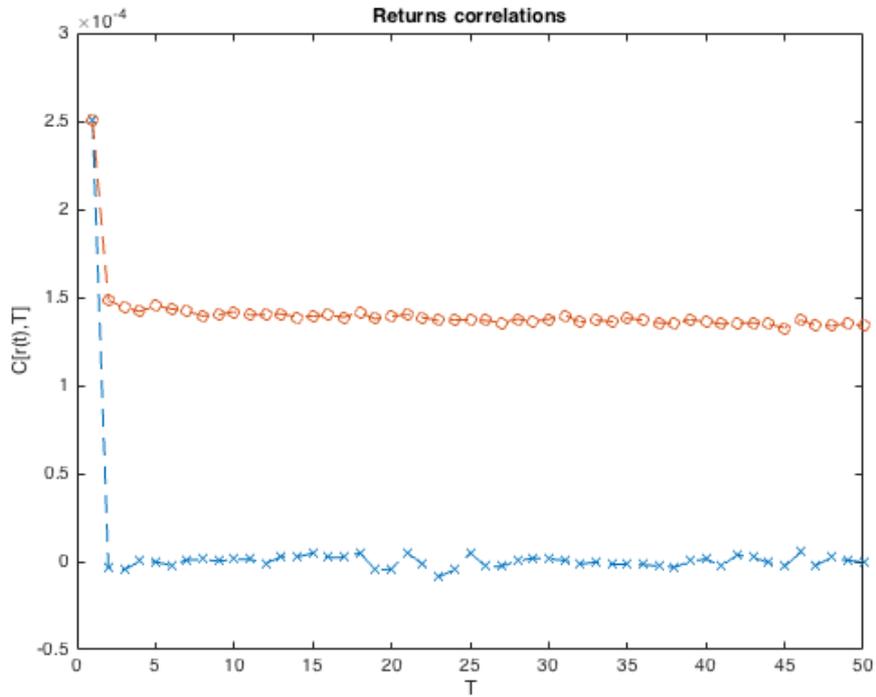

Figure 10: Correlations of $r(t,1)$ and $|r(r,1)|$

## 5. Reconstructing the process

In order to reconstruct the process, we perform the following steps in order:

1. Get the stock value prices $p(t)$
2. Calculate the 1-day returns $r(t,1)$
3. Calculate the average $\langle r(t,1) \rangle$ and standard deviation $\sigma$
4. Choose an N-symbol alphabet $\Sigma$, ideally $N \geq 3$ and odd
5. Determine the coded sequence $S$ with $s_i \in \Sigma$
6. Split the sequence $S = \{s_1, \ldots, s_N\}$ in half into $S_1$ and $S_2$ such that $S = S_1 \cup S_2$, $S_1 = \{s_1, \ldots, s_m\}$ and $S_2 = \{s_{m+1}, \ldots, s_N\}$ with $m = \left\lfloor \frac{N}{2} \right\rfloor$ where $\lfloor x \rfloor$ is the floor function

7. Use the first half $S_1$ to calculate the transition matrix $P$
8. Use the second half $S_2$ and the transition matrix $P$ to forecast $S_2^*$

After obtaining the forecast, we need something to compare it to. Simultaneously, we construct a randomly generated sequence $R$ with the same length as $S_2^*$ but with each element $r_i \in \Sigma$ generated at random from a uniform distribution. The error is calculated for both sequences $S_2^*$ and $R$ according to eq. (8) and we can compare how using the transition matrix $P$ performs against a completely random process.

The process to determine both $S_2^*$ and $R$ is as follows.
1. Read the value $s_i \in S_2$
2. Extract the corresponding column $P(*|s_i)$ from the transition matrix
3. Throw a random number $\varepsilon$ from the uniform distribution $u(0,1)$
4. Set $s_i^* = \min_x \{x \in \Sigma : \varepsilon \leq P(x|s_i)\}$
5. Set $r_i$ to a random element of $\Sigma$

Taking the total of 500 simulations as represented in figure 11, we were able to obtain for a 3-symbol 1-length Markov process an error of $e_1$ and for a completely random process an error of $e_2$.

$$e_1 = 0.35106$$
$$e_2 = 0.66646$$

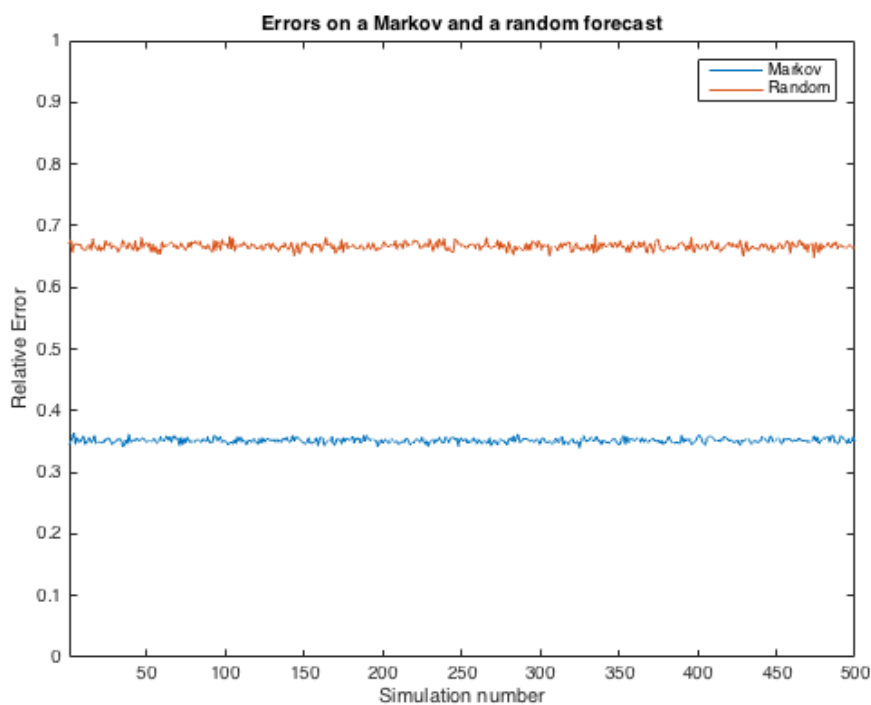

Figure 11: Comparison of the error obtaining in using the Markov process vs. a random process

The transition matrix obtained was:

$$P = \begin{bmatrix} 0.1391 & 0.0832 & 0.1308 \\ 0.7202 & 0.8190 & 0.7425 \\ 0.1407 & 0.0978 & 0.1267 \end{bmatrix}$$

Results show that it's much more likely for the price to remain constant within one standard deviation than to go either up or down by more than one standard deviation.

We used the first half of the sequence $S_1$ to construct the transition matrix which helped us in forecasting $S_2^*$, that is <u>the past predicting the future</u>. We could also have inverted the process in order to obtain <u>the future predicting the past</u>. This is easily achievable by using the second half sequence $S_2$ to construct the transition matrix and proceed exactly as before to obtain $S_1^*$.

Another thing that we can do is reverse the process in order to obtain the predicted prices $p^*(t)$ from the returns forecast sequence $S_2^*$. The results can be observed in figure 12. We can observe that, even though the forecast (red) may not align with the actual prices (blue), it's almost always performs better than a completely random forecast (yellow), especially in the long run.

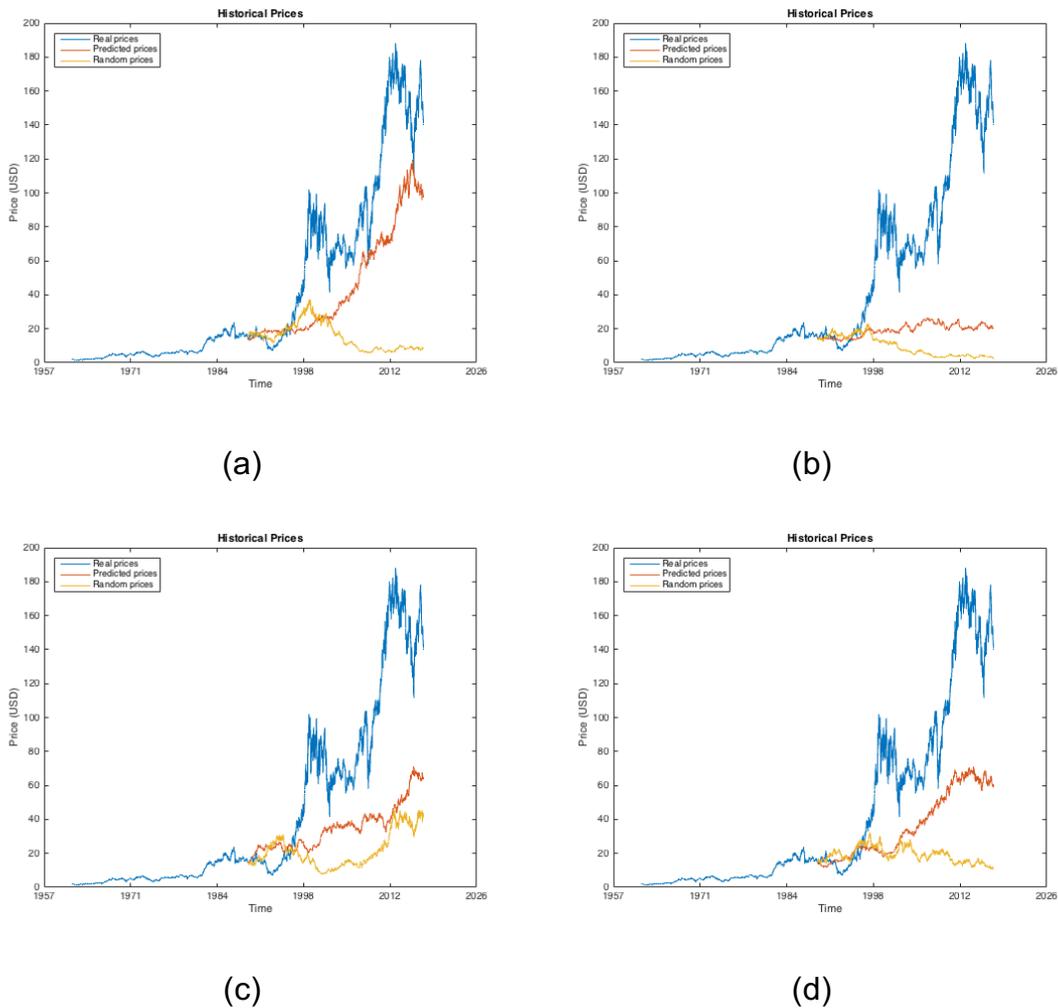

Figure 12: Different runs of the reverse process to reconstruct the prices from the forecasted returns

Since the results can vary greatly with each run, we can do a Monte Carlo simulation to observe the underlying tendency as in figure 13. For a total of 500 simulations, we observe clearly that the Markov forecast is closer to the actual price than the random one and the difference grows larger in time.

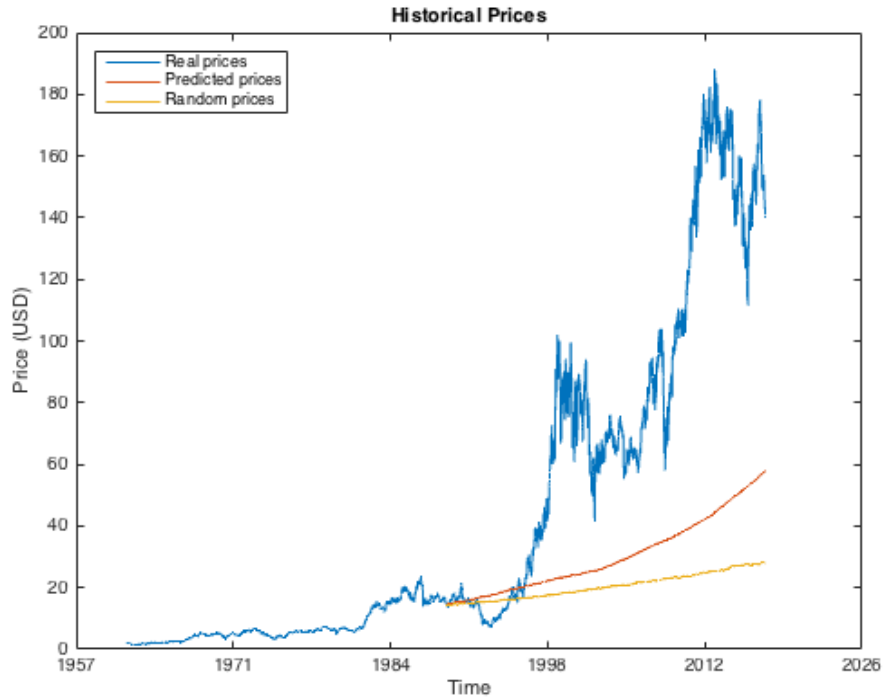

Figure 13: A Monte Carlo simulation for the reverse Markov process

**5.1. Increasing N and K**

We can use a more generalized algorithm to perform the previous Markov reconstruction using a N-symbol and K-length Markov chain. We did a computation for a 5-symbol alphabet with K ranging from 2 to 8 with 10 simulations each and the average errors according to (8) are presented in figure 14. We can observe once again that the error with a randomly generated sequence is greater than using the Markov chain process. We also observe that the error initially decreases with increased lengths for the chain, but for $K \geq 5$ it increases again. One explanation might be that for $K \leq 4$ there are enough K-length chains in the data, but longer chains might be rarer.

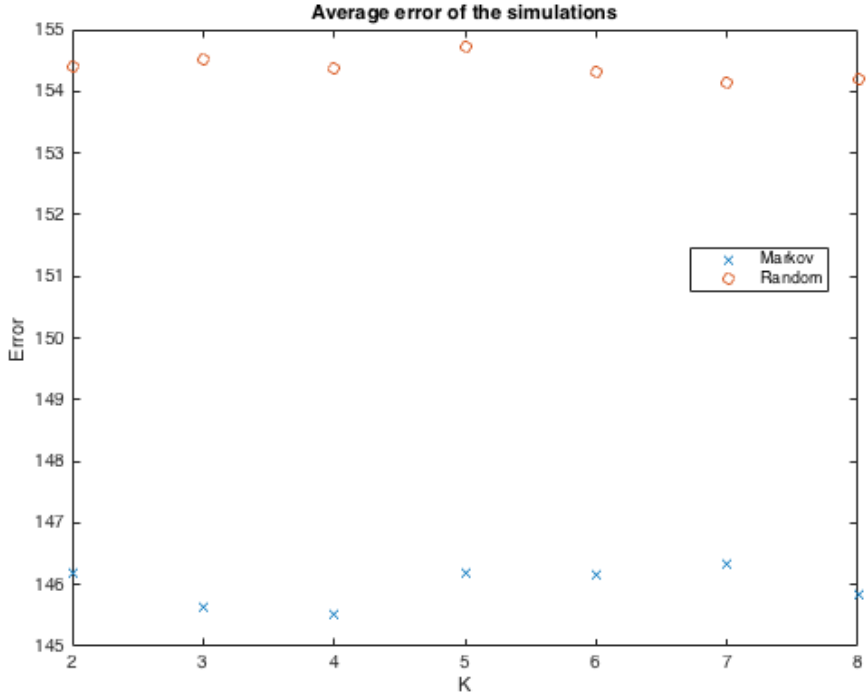

Figure 14: Average error for K-length Markov chain simulations

**5.2. Reverse the Markov coding process**

The results in figures 12 and 13 were obtained by doing a reverse coding process in order to obtain the price values $p^*(t)$ from the forecasted sequence $S^*$. We can find the predicted returns $r^*(t)$ by applying eq. (18) to each $s_i^* \in S^*$.

$$r_i^* = \langle r(t,1) \rangle + \frac{2\sigma_r}{N-1} s_i^* \qquad (18)$$

Where $\langle r(t,1) \rangle$ and $\sigma_r$ are the mean and standard deviation of the 1-day returns $r(t,1)$, $N$ is the number of symbols in the alphabet and $s_i^* \in \Sigma$. Then, we simply need to use eq. (19) in order to obtain the predicted prices from the predicted returns.

$$p_i^* = p_{i-1}^* e^{r_i^*} \tag{19}$$

## 6. The algorithm for the transition matrix

The algorithm we used is general and can be used to calculate the transition tensor for any N-symbol and K-length Markov chain. Let's start with the simplest 3-symbol and 1-length case with the transition matrix $P$. We have the sequence $S = \{s_1, ..., s_N\}$ which contains the coded returns $r(t)$ for the prices $p(t)$ using an alphabet $\Sigma = \{-1, 0, 1\}$ such that $s_i \in \Sigma$.

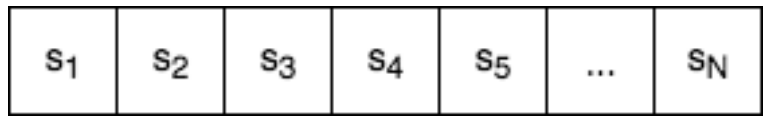

Figure 15: Graphical representation of an array

We also need to define what an array data structure is. More commonly known as simply an *array*, it's a collection of elements each of which identified by an array index or key. The simplest one-dimensional array is similar to a vector. We can use an array to store our sequence $S$, for example, and we reference each element $s_i$ by its index number $i$ as in figure 15. If $A$ is our array, then $A[m] = s_m$. An array can also be referenced by a keyword, e.g. $A[one] = s_1$, or we can have a multidimensional array in which indexes are also arrays. We are going to make use of the latter to build our transition tensor $T$.

The algorithm simply counts the number of times a given sequence occurs and divide that by the total number of sequences counted. In order to achieve that, we use arrays in which the indexes are the sequences themselves and add one unit every time they come up.

We're going to need two multidimensional arrays $A$ and $B$ and we'll store the frequency of occurrence for each unique sequence in $A$ and the total sequence count in $B$. We need to read subsets $X_u$ of $S$ with a length of $K+1$ one at a time in sequence. In this case, we have $X_1 = \{s_1, s_2\}$, $X_2 = \{s_2, s_3\}$, ..., $X_{n-1} = \{s_{n-1}, s_n\}$. We also need the subset $Y_u$ of $X_u$ which is simply $X_u$ with the last element removed. For example, $Y_1 = \{s_1\}$, $Y_2 = \{s_2\}$, ..., $Y_{n-1} = \{s_{n-1}\}$.

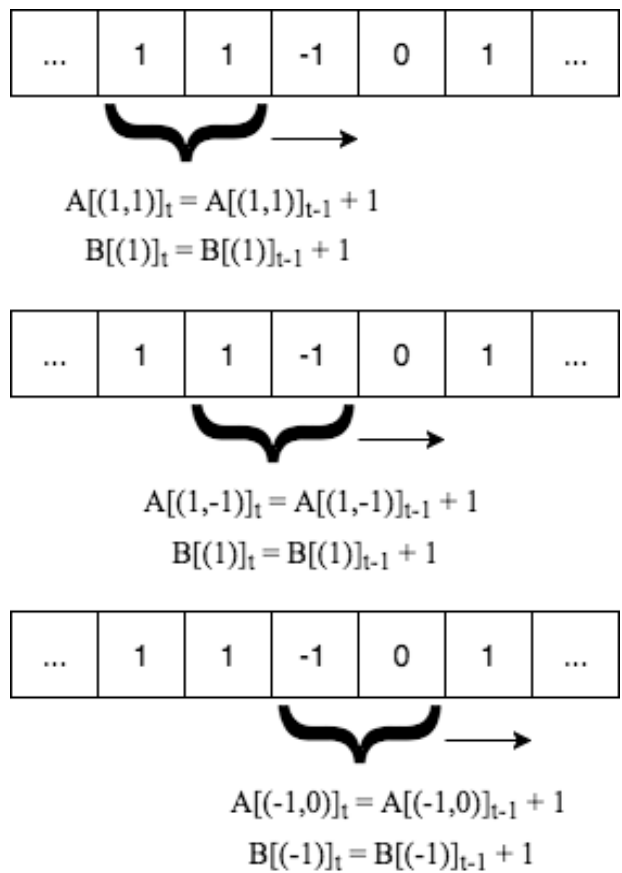

Figure 16: Graphical representation of the algorithm

Then, we can use $X_u$ and $Y_u$ as indexes for $A$ and $B$, respectively, and every time they occur we add one unit to its current value. At each occurrence of $\chi_u$, we have $A[X_u]_t = A[X_u]_{t-1} + 1$ and $B[Y_u]_t = B[Y_u]_{t-1} + 1$ as in figure 16.

Let $\#\{w\}$ be the number of times the element $w$ of $\Sigma$ appears in the sequence $S$, we have that $A$ is equivalent to an $N \times N$ matrix and $B$ to an $N \times 1$ vector:

$$A = \begin{bmatrix} \#\{-1,-1\} & \#\{0,-1\} & \#\{1,-1\} \\ \#\{-1,0\} & \#\{0,0\} & \#\{1,0\} \\ \#\{-1,1\} & \#\{0,1\} & \#\{1,1\} \end{bmatrix}$$

$$B = \begin{bmatrix} \#\{-1\} \\ \#\{0\} \\ \#\{1\} \end{bmatrix}$$

To build our transition matrix $P$, we simply need to extract each $N \times 1$ column from $A$ and divide each element by the respective element of $B$:

$$P = \begin{bmatrix} P(-1|-1) & P(-1|0) & P(-1|1) \\ P(0|-1) & P(0|0) & P(0|1) \\ P(1|-1) & P(1|0) & P(1|1) \end{bmatrix} = \begin{bmatrix} \frac{\#\{-1,-1\}}{\#\{-1\}} & \frac{\#\{0,-1\}}{\#\{0\}} & \frac{\#\{1,-1\}}{\#\{1\}} \\ \frac{\#\{-1,0\}}{\#\{-1\}} & \frac{\#\{0,-1\}}{\#\{0\}} & \frac{\#\{1,-1\}}{\#\{1\}} \\ \frac{\#\{-1,1\}}{\#\{-1\}} & \frac{\#\{0,-1\}}{\#\{0\}} & \frac{\#\{1,-1\}}{\#\{1\}} \end{bmatrix}$$

And this can be generalized to build the transition tensor $T$ for any N-symbol K-length Markov chain.

## 7. Conclusion

The financial markets are apparently random as we have seen in the similarity between figures 1 and 5 where the former shows a completely random and the latter a real stock market.

We have shown evidence for some of the stylized facts about the financial markets such as the existence of memory-like phenomena in price volatility in the short term from the accumulated volatility in figure 7 and the correlations in figure 10, a power-law behavior and non-linear dependencies on the returns in figure 8. This comes in contrast with the EMH.

Given the observed facts, we modelled the market using Markov chains as they are memoryless and dependent only on the current state of affairs. That is, the bulk of the fluctuations appear to be a short-term process with a small long-memory component. These might, in turn, be responsible for the larger fluctuations in returns.

We have shown, as seen in figures 11 through 14 that the reconstruction of the market process from the Markov transition matrix is, at the very least, always better than the completely random process when comparing both to the actual data.

If we're to believe Warren Buffet in that the emotions control de short-run, then we are correct in assuming that given the present state of affairs people ought to behave in a certain way. That is, people behave predictably with some probability given the *status quo*. Thus, the implemented Markov chain is a good tool in that it takes the present state and weighs it against similar states in the past in order to obtain the transition probabilities for all possible outcomes.

We have seen that for every day, there is a certain probability of the price of an asset to remain the same, go up or go down. The probabilities for each event are conditional on past behavior, assuming that the next price movement depends only on the present state of affairs where the past weighs in on the transition probabilities.

Thus, we've built an algorithm that takes all of this into account and is as general as it can be. In its simplest way, we can say that the price remains the same if it doesn't vary more than one standard deviation either up or down. If it exceeds this, then we consider an upwards or downwards movement, respectively. Our algorithm takes into account that we might want to refine this interval, instead of diving it in 3 possibilities, we can do it for $N \geq 3$ possible outcomes.

After we concluded our analysis using a Markov chain of length 1, we decided to generalize that for any K-length sequence. Thus, instead of taking into account only the $n-1$ day when building the transition matrices, we consider the whole sequence of events in the $n-K$ days before.

As a last comment, we can interpret the transition matrices as a measure of market sentiment, the overall attitude of investors towards a particular asset. If people follow their emotions and there is a higher probability of the price to go up, then we can say market sentiment is bearish. Otherwise, we'd say market sentiment is bullish.

There are some market sentiment indicators available such as the CBOE Volatility Index (VIX), 52-week High/Low Sentiment Ratio, Bullish Percentage, 50-day moving average and 200-day moving average. An improvement of the current work could be to incorporate some of these measures into the algorithm.

It becomes important as the agents operating in the markets are fundamentally human and not emotionless highly intelligent beings or *econs*, according to the 2017 Nobel laureate Robert Thaler. As he explains in reference (Thaler, 2015), there is no reason why certain dates, e.g. anniversaries, football matches, etc., should matter more than others in the point of view of an *econ* but we know they do and it reflects on the markets, e.g. as a significant decline after football losses (Edmans, Garcia, & Norli, 2007).

The model presented here is available in GitHub for scrutiny and improvement and ready to be applied to any data source. A general model for any N-symbol and K-length Markov chain was developed as an improvement to the work presented in reference (Vilela Mendes, Lima, & Araújo, 2002). The same process used to obtain the results presented here have been replicated for Facebook (FB) and Google (GOOG) and are presented in the appendix.

## 9. Appendix

Some of the same techniques from this work are replicated here for different companies: Facebook (FB) on the left and Google (GOOG) on the right.

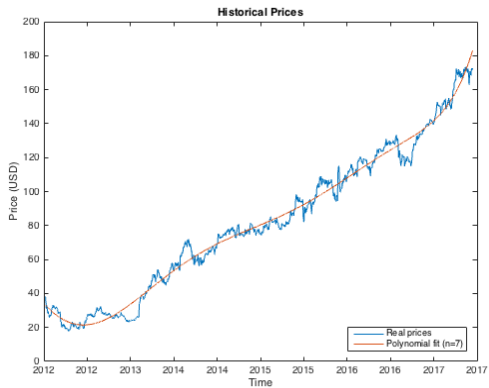 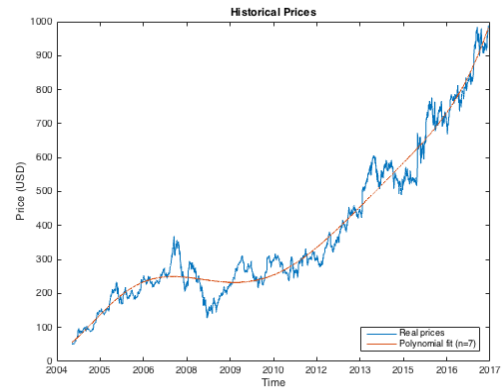

(a)                  (b)

Figure A.1: (a) Facebook's and (b) Google's stock daily closing price over time and polynomial fit

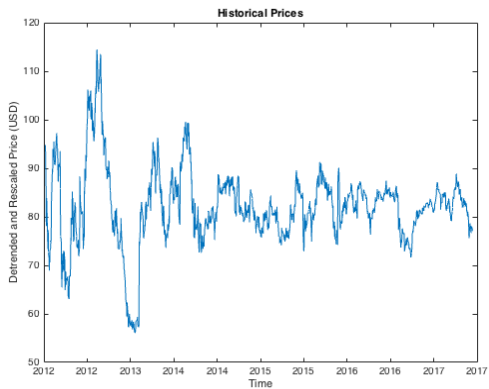 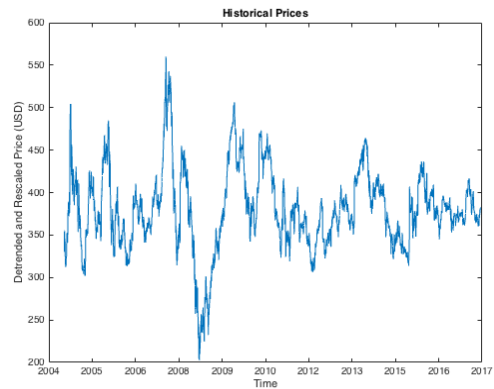

(a)                  (b)

Figure A.2: Rescaled prices for (a) Facebook and (b) Google

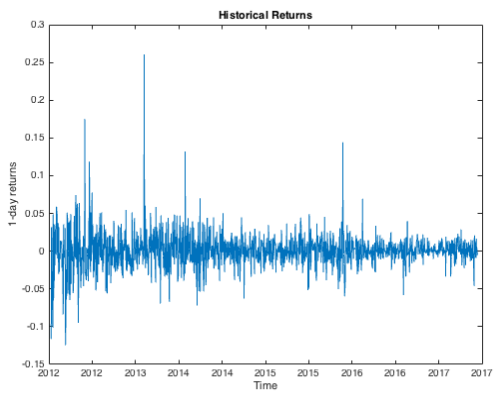 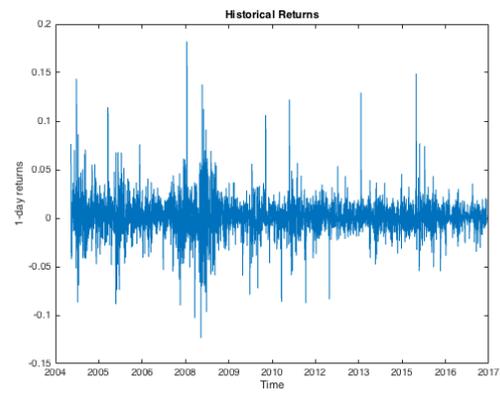

(a)                       (b)

Figure A.3: Historical returns for (a) Facebook and (b) Google

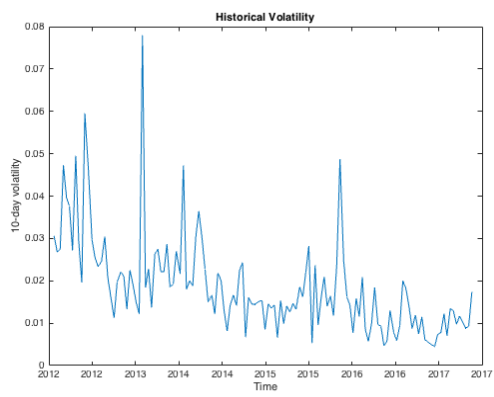 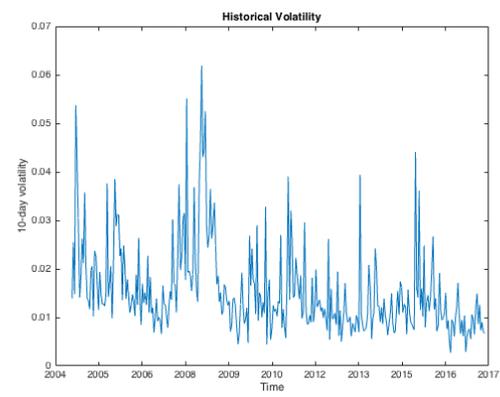

(a)                       (b)

Figure A.4: Historical volatility for (a) Facebook and (b) Google

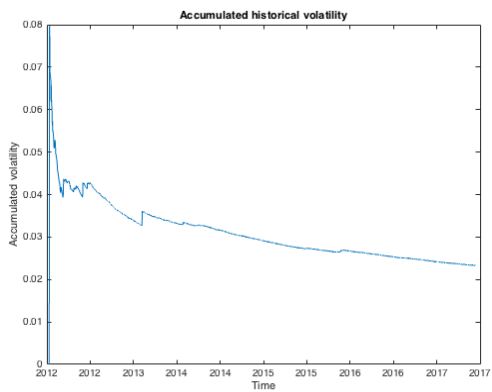 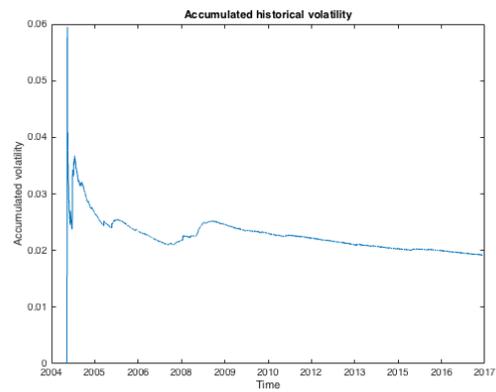

(a)                          (b)

Figure A.5: Accumulated volatility for (a) Facebook and (b) Google

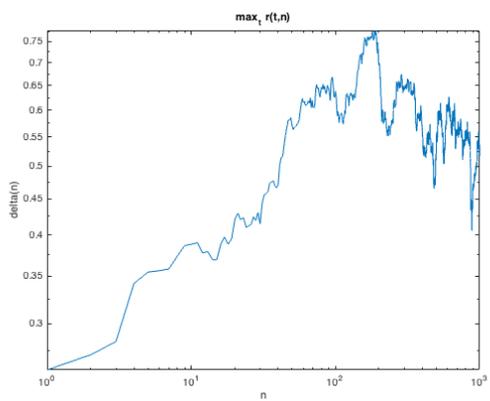 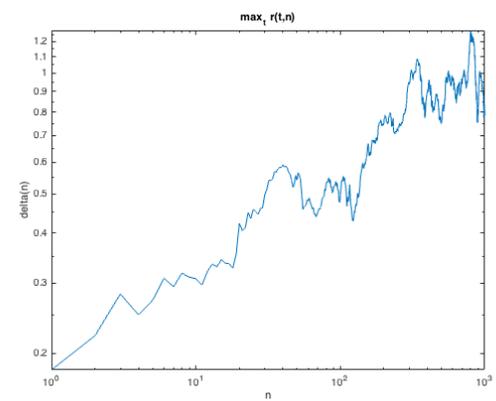

(a)                          (b)

Figure A.6: The plots of the maximum $\delta(n)$ with $n$ from 1 to 1 000 for (a) Facebook and (b) Google

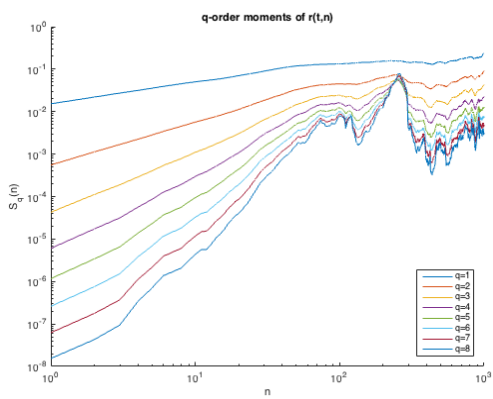
(a)

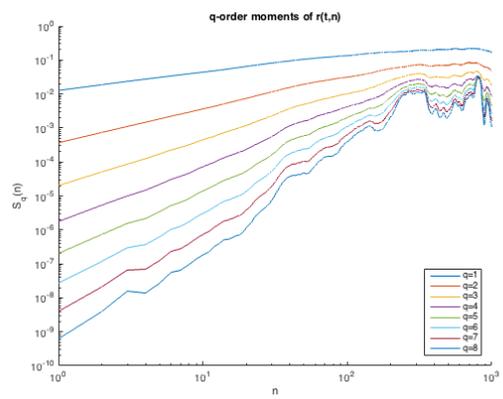
(b)

Figure A.7: The plots of the moments $S_q(n)$ with $q$ from 1 to 8 from top to bottom for (a) Facebook and (b) Google

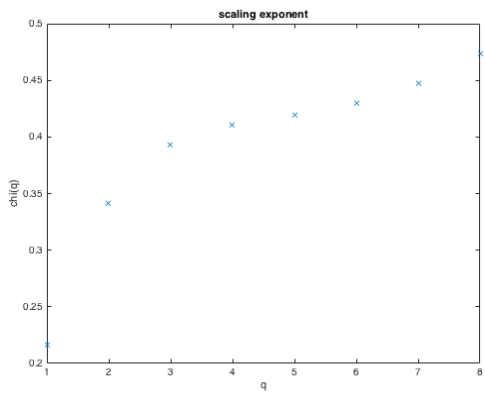
(a)

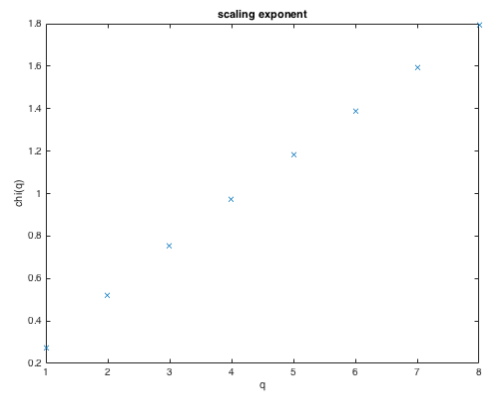
(b)

Figure A.7: The scaling exponent $\chi(q)$ for (a) Facebook and (b) Google

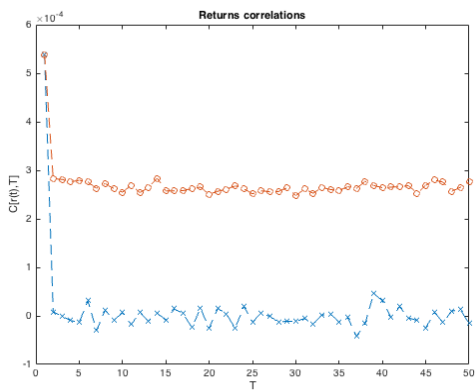 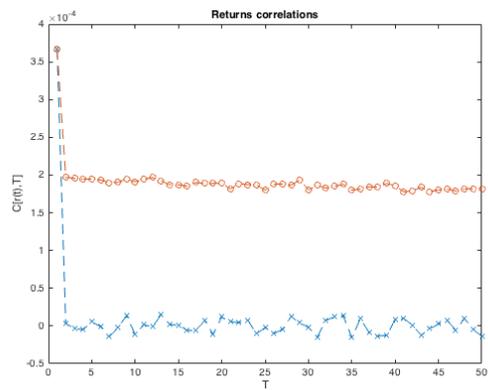

(a)                          (b)

Figure A.8: Correlations of $r(t,1)$ and $|r(r,1)|$ for (a) Facebook and (b) Google

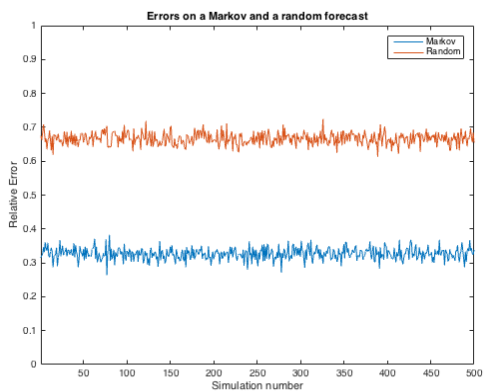 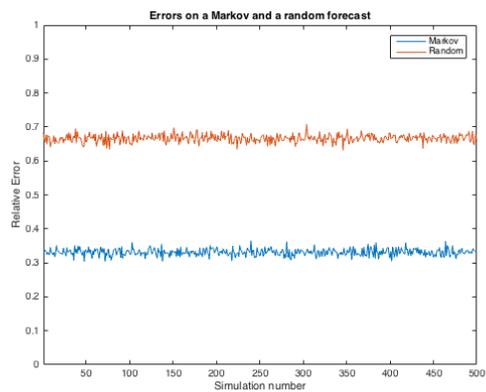

(a)                          (b)

Figure A.9: Comparison of the error obtaining in using the Markov process vs. a random process for (a) Facebook and (b) Google

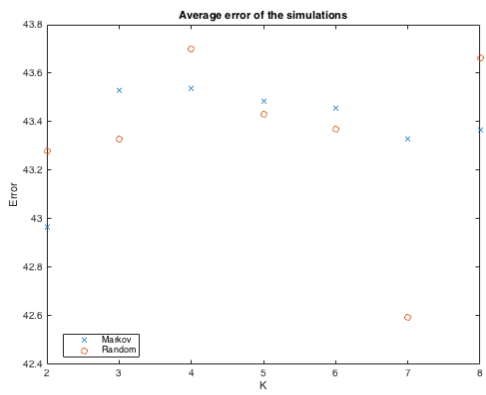 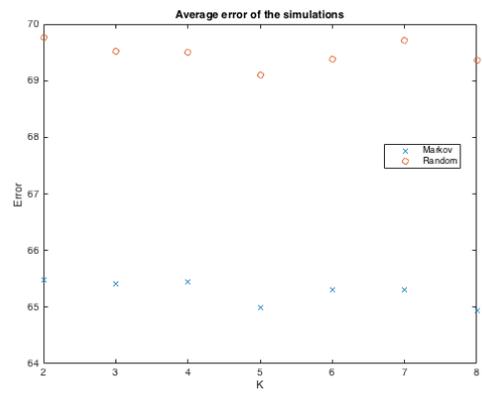

(a) (b)

Figure A.10: Average error for K-length Markov chain simulations for (a) Facebook and (b) Google

44